\documentclass[man,11pt]{apa7}
\usepackage{amssymb}
\usepackage{framed}
\usepackage[american]{babel}
\usepackage{csquotes}
\usepackage{fancyhdr}

\usepackage{enumitem}
\setlist{nosep}
\usepackage[style=apa,backend=biber]{biblatex}
\title{The Technology of Outrage: Bias in Artificial Intelligence}
\shorttitle{Technology of Outrage}
\leftheader{Technology of Outrage}
\authorsnames[1,1,2]{Will Bridewell,Paul F. Bello, Selmer Bringsjord}
\authorsaffiliations{{U.S. Naval Research Laboratory}, {Rensselaer Polytechnic Institute}}
\abstract{Artificial intelligence and machine learning are increasingly used to offload decision making from people. In the past, one of the rationales for this replacement was that machines, unlike people, can be fair and unbiased. Evidence suggests otherwise. We begin by entertaining the ideas that algorithms can replace people and that algorithms cannot be biased. Taken as axioms, these statements quickly lead to absurdity. Spurred on by this result, we investigate the slogans more closely and identify equivocation surrounding the word `bias.' We diagnose three forms of outrage---intellectual, moral, and political---that are at play when people react emotionally to algorithmic bias. Then we suggest three practical approaches to addressing bias that the AI community could take, which include clarifying the language around bias, developing new auditing methods for intelligent systems, and building certain capabilities into these systems. We conclude by offering a moral regarding the conversations about algorithmic bias that may transfer to other areas of artificial intelligence.}
\date{25-JAN-2022} 

\begin{document}

\maketitle

\section{Introduction}

Everywhere we turn, we are confronted with stories about how artificial intelligence (AI) can be used to streamline decision-making in business, the government, and military defense. In the era of rule-based expert systems, there was a belief that rationality could be automated. More recently, the goal has been to derisk decision-making by learning from statistical regularities in data. Regardless of the approach, there has been a hope that when deployed these tools could reduce the influence of human biases when choosing courses of action. The reality shown by an increased reliance on machines is that bias is unavoidable. Therefore, promoting fair and impartial outcomes requires vigilance even when supposedly unbiased algorithms are used. This result should come as no surprise, because if we take as axiomatic that machines can replace people and that machines cannot be biased, we are forced into an inference that is patently absurd. Nevertheless, public discussions about algorithmic bias can become heated. Our intention is to use this absurdity as a basis for dissecting bias and the source of the associated outrage in the context of AI.

One of the central tenets of AI, at least among those who think that artificial human-like intelligence is possible, can be sloganized as ``People Are Algorithms.''\footnote{We follow folk convention where \textit{algorithm} refers to both the mathematical abstraction that is a set of instructions for executing a procedure and its corresponding implementation within a physical system.} This claim extends back to the origins of the field and is mirrored in the thought-is-computation metaphor embraced within cognitive science \parencite{Newell1976}, philosophy \parencite{Putnam1975} and neuroscience \parencite{McCulloch1943,McCulloch1960}. Coupled with a pervasive undertone of Cartesian (mind-body) dualism that identifies the person with the mind, we arrive at the slogan by the following syllogism.
\begin{quote}
    \textbf{Syllogism 1}\\
    People are minds.\\
    Minds are algorithms.\\
    $\therefore$ People are algorithms.
\end{quote}
Although it is possible to preserve the goals of AI while rejecting the first premise, rejecting the second would require the field to abandon its primary conceit and accept a somewhat deflated goal of writing programs to address a proper subset of computationally challenging problems. 

As AI-enabled systems are deployed, public concerns have arisen that not only do these systems exhibit human-like behavior and inform decisions ordinarily made by humans, but they also reflect all-too-human prejudices. And the public is right to be concerned, whether in the case of Microsoft's chatbot Tay \parencite{Wolf2017}, the classification of images in Google Photos \parencite{Dougherty2015,Vincent2018}, Facebook's automated labeling of videos \parencite{Mac2021}, or r\'esum\'e screeners that prioritize applicants named Jared \parencite{Gershgorn2018}. In response, a vocal community of computer scientists have pushed back on these concerns with their own slogan, ``Algorithms Cannot Be Biased.'' The purpose of this claim is not to deny the well-documented cases, but to preserve the formal perspective that algorithms are mathematical abstractions that, as such, cannot inherently encode prejudices. With some expository license, we can summarize this view in another syllogism.
\begin{quote}
    \textbf{Syllogism 2}\\
    Algorithms are mathematical abstractions.\\
    Mathematical abstractions cannot be biased.\\
    $\therefore$ Algorithms cannot be biased.
\end{quote}
To be clear, no one disputes that designing an algorithm (or a mathematical equation, formula, etc.) that produces biased output is trivial. The public conversation is concerned with whether biased output results unintentionally from standard AI algorithms or systems and where the source of bias lies. 

Agreement with the conclusions of both syllogisms is not uncommon among AI researchers. As an example, Pedro \textcite{Domingos2015} discusses what he calls the \textit{master algorithm}: a hypothetical, single algorithm that subsumes all particular algorithms for machine learning. To argue that such an algorithm exists, Domingos states that human brains execute the master algorithm (the precise nature of which, he readily concedes, is currently unknown). So, we are ourselves confirmation that the master algorithm exists. The first slogan follows immediately from Domingos' assertion: we are algorithms. More recently, the same author has written:
\begin{quote}
[M]achine-learning algorithms, like pretty much all algorithms you find in computer-science textbooks, are essentially just complex mathematical formulas that know nothing about race, gender or socioeconomic status. They can't be racist or sexist any more than the formula $y = ax + b$ can. \parencite[][para. 5]{Domingos2020}
\end{quote}
This quote is a particular elaboration of the second slogan: algorithms cannot be biased.  

Now, something interesting happens when both syllogisms are taken together.\footnote{The sense of `algorithms' in both premises includes those currently in use as machine-learning algorithms.}
\begin{quote}
    \textbf{Syllogism 3}\\
    People are algorithms.\\
    Algorithms cannot be biased.\\
    $\therefore$ People cannot be biased.
\end{quote}
We assume that readers will see the immediate absurdity of the conclusion, evidence against it being readily available. Regardless, the conclusion is sanctioned by joint belief in the premises that we established as common if not uncontroversial among AI researchers. Further, Syllogism 3 and its absurdity hold for any specialized interpretation of bias, such as racism, sexism, and ageism.

In the remainder of the paper, we use the need to neutralize the absurdity-producing Syllogism 3 as the chief engine for moving forward. Specifically, we explore the idea that people may not be algorithms, or at least not \textit{solely} algorithms. Then we take a closer look at bias, whose various interpretations sometimes tangle up discussions. After scrutinizing the individual premises, we rewrite Syllogism 3 in a way that lets people eat their cake and have it, too. Shifting focus to the sometimes-heated disagreements surrounding algorithmic bias, we claim that arguments framed in terms of the slogans mask deeper concerns about personhood and dehumanization. We conclude by describing ways to soothe these concerns and point out that similar ones may arise when other aspects of AI are brought to the fore.  

\section{People Are Algorithms?}

The first premise in Syllogism 3 is that people are algorithms. Many readers outside the fields of AI and cognitive science might find themselves strongly inclined to reject this claim and treat the syllogism's absurdity as properly addressed. After all, so many of our ideas about personhood seem immune to mathematical representation: 
\begin{itemize}
    \item people have bodies and move within the world; 
    \item people have vibrant, full, subjective experiences; 
    \item people have the freedom to choose how to act, a freedom informed by questions of right and wrong that have created entire cultures;
\end{itemize}
and the list continues. However, there are other readers in the laity who not only entertain but also champion such ideas as that we are living in a simulation, that we can upload our minds to a computer and achieve digital immortality, and that we can build artificial agents indistinguishable from humans. These beliefs appear to agree with some version of the premise and appear as frequently in popular culture as in academic discourse.

How could someone believe that what makes us people is some particular algorithm, an abstraction? The first syllogism provides one potential answer. In his \textit{Meditations on First Philosophy}, \textcite{Descartes2008} argued that we are, at our indubitable foundations, thinking things and that these thinking things are formed of a substance separate from matter. On this view, which was part and parcel of scientific thought after Descartes, \textit{a person is a mind}; that the only minds we know of appear in a synthesis of mind and body is mere happenstance. In the same period \textcite{Hobbes1839} published his \textit{De Corpore}, in which he posited that reasoning is computation. His perspective was reflected in Boole's \parencite*{Boole1854} \textit{Laws of Thought}, in Frege's \parencite*{Frege1980} \textit{Foundations of Arithmetic}, and hence in the exploration of formal logics throughout the 20th century. These developments in turn influenced the earliest thinkers in cognitive science, where minds were seen to be information processors much like the newest computers, and in artificial intelligence, where these computers were seen to finally be powerful enough to mechanically execute supposed rules of thought \parencite{Turing1969}. From their inceptions, the idea that \textit{the mind is an algorithm} has been a central tenet of these fields, and the question of whether people are algorithms has been a perennial point of discussion.

\begin{framed}
\begin{quote}
    The ultimate goal of AI (which we are very far from achieving) is to build a person, or, more humbly, an animal. \parencite[][p. 7]{Charniak1985}
\end{quote}
\begin{quote}
    The human cognitive mind/brain is a computational device (computer); hence, the human cognitive capacities consist, to a large extent, of a system of computational capacities. \parencite[][p. 98]{Eckardt1995}
\end{quote}
\end{framed}

Regardless, as we have indicated, our lived experiences stand at odds with the impersonal proclamation that we are the fruit of some abstract mathematical recipe for turning sensory information into action. Consequently, one might reject either that personhood can be isolated to the mental realm or that minds are essentially computer programs. For instance, advocates of embodied and phenomenological approaches to cognition \parencite[e.g.,][]{Gallagher2005,Zahavi2021} reject the Cartesian divide between mind and matter. Instead, they claim that cognition is inseparable from physical bodies and that the proclivity to treat mental activity as disembodied hinders progress in understanding cognition. The embedded perspective on cognitive science often co-occurs with calls to reject purely computational views of cognition that rely on the manipulation of symbolic representations. However, one can identify persons with minds while also understanding a mind to be something different from or more than an algorithm. To this point, the mind may be an algorithm plus experiential history. That is, while the algorithm executes, making inferences or whatever else may be in its purview, non-algorithmic mental processes may alter the operation of the supposed algorithm or override its output. Alternatively, if the concept of an algorithm is tied to the limits of Turing computability, then one could appeal to ideas of hypercomputation \parencite{Bringsjord2004} to reject the premise that the mind is an algorithm while saving cognitive science's commitment to information processing.

Although one can reject that people are algorithms, this position is easier to take in cognitive science than in artificial intelligence. Within AI there are a variety of weak and strong views, but these tend to take the following forms.\footnote{Early conceptions of weak and strong AI raised the metaphysical question of whether a computational model of thinking was actually thinking or only simulating thought. Over the decades, the distinction has been repurposed, and currently, strong AI is associated with artificial general intelligence and weak AI emphasizes isolated, although potentially challenging, activities. The definitions given here are an attempt to distill the claims that would motivate these two perspectives.}
\begin{quote}
    \textbf{Weak AI:} a sophisticated enough computer program can produce any single intelligent behavior that a human can.\\
    \textbf{Strong AI:} a single, sophisticated enough computer program can produce all intelligent behaviors that a human can (possibly limited to those that do not require physical action).
\end{quote}
Proponents of strong AI cannot reject the premise under discussion without directly contradicting their core claim unless they define persons as a combination of strong AI with some other (measurable) property that is unnecessary for reproducing all human behaviors. Interestingly, the same is true for proponents of weak AI because of a commitment to the claim that \textit{any} single thing a person can do, a computer can also do. So, even if machines can defeat human experts at various games, there are legions of other activities left to mimic. Whether these have an algorithmic solution is an open scientific question, and if even one of them does not, then the weak AI position fails. Therefore, to the extent that one is an optimist about the research program of AI, it makes sense to turn elsewhere as a way out of the absurdity that we revealed.

\section{Can Algorithms Be Biased?}
\label{sec:bias}
The second premise in Syllogism 3 is that algorithms cannot be biased. As with the first one, this assertion may seem strange to readers outside of computer science and AI. After all, the introduction listed multiple examples of programs that exhibit measurable, biased output. However, the claim interprets both `bias' and `algorithm' in ways that, while reliant on technical definitions, shift the responsibility for any bias to the people who develop, validate, and deploy AI systems that affect human lives. Friedman and Nissenbaum \parencite*{Friedman1996} provide a taxonomy of the sources of bias in technology, and here we specifically discuss forms of pre-existing and technical bias that are relevant to machine-learning algorithms, which currently attract considerable attention. By scoping our concerns specifically to learning algorithms, we are positioned to refine Syllogism 3 in a manner that avoids absurdity, but perhaps not impracticality. 

\subsection{Inductive Bias}
When describing machine-learning algorithms, one form of bias is unavoidable: inductive bias. Consider a basic example where there is a dataset with an independent variable $x$ and a dependent variable $y$. One model for the relationship between these variables is $y = Ax + B$, where $A$ and $B$ are    parameters. This is a linear equation, so any predictions that the model might make will fall on the straight line defined by the values of the parameters. If the relationship between $x$ and $y$ is linear, then the model can be parameterized with as few as two points. However, if the true relationship is a nonlinear polynomial (e.g., $y = x^2$), then the model's inductive bias will impair its accuracy regardless of the amount of data available. 

Importantly, inductive bias is not only about selecting the correct structure for a model but also about a variety of other assumptions that enable machine learning \parencite{Dotan2021,Wolpert1997}. For instance, there may be constraints on the values that the parameters can take, so for interpretable models where the parameters are associated with measurable quantities in the world, their ranges may be determined by laws of physics or expert knowledge \parencite{bridewell-2k8,Bridewell2006}. Likewise, there can be explicit constraints on model structure or viability that rule out certain solutions (e.g., if the model makes predictions that violate well-established cause-effect relationships; \cite{Pazzani2001,Bridewell2004}). Additionally, objective, or loss, functions guide learning algorithms through the space of valid models, biasing their search in ways that can lead to locally optimal solutions at the expense of generalizability to new examples. Finally, the algorithm that updates models may itself include biases that make certain solutions impossible. Often these inductive biases are introduced intentionally by people to guide the learning process away from unpromising, infeasible, or incomprehensible solutions. However, unintentional biases are also common. For instance, when neural network researchers report on approaches that work well for one kind of data versus another, they are noticing the effects of inductive biases built into particular network architectures, optimization algorithms, and loss functions, not all of which may be well understood. 

\subsection{Sample Bias}
In contrast to inductive bias, which is a property of the learning algorithm, sampling bias is a property of the data used for training. Examples of poor sampling include using Twitter content or Victorian era literature to train a model of natural language generation or using images of only light-skinned people to train a model of facial recognition. Technically speaking, sampling bias is not algorithmic bias. Active learning provides a caveat to this view because algorithms in this paradigm can modify or guide the contents of their training data \parencite{Settles2009,Cohn1996}. Use of active learning is rare in practice, which means that determining whether the sampling bias is responsible for a particular model's biased performance requires access to the training data or meaningful statistics that describe it.

Faced with sampling bias, a common suggestion is to balance the data or to make it more reflective of the population. These techniques may be useful, but they are not panaceas due to the presence of unknown relationships among variables. For instance, suppose that a dataset was found to be unbalanced in terms of economic class, with few entries reflecting low-income households. To balance that variable in the dataset, whether to make it more reflective of the population or to over-represent certain categories as a way to increase the importance of low-income households during learning, a sample of convenience may be drawn from a single geographic region. As a consequence, the data, although adjusted on that one feature, would provide a limited picture of low-income households in general, one that would be specific to the sampled location. From these data, machine-learning algorithms would produce a model that makes predictions about income levels that would fail when applied to other locations. Arbitrary relationships are hard to avoid without expert care in constructing a data set. This difficulty is compounded when learning algorithms incorporate methods for feature engineering because otherwise undetected and incidental relationships among input features can be amplified in the final model. Often the only solution is to extensively evaluate the learned model and to interleave cycles of model development and data curation.

\subsection{Systemic Bias}
\label{sec:systemic-bias}
Inductive and sampling bias may be root causes of broad partiality, but they are not what typically catches the public's eye when algorithmic bias becomes newsworthy. Instead, the stories refer to bias exacerbated by trained and deployed software, such as Northpointe COMPAS \parencite{CorbettDavies2016}, Equivant's risk assessment tool for criminal recidivism. Broadly, the emphasis is on aspects of social or systemic biases that are reflected in or strengthened by AI software or other algorithms. These biases may persist even when care is taken to ensure social-value-neutral inductive biases and to avoid sampling bias. One way that they can occur is when the only data available reflect systemic social biases. In this case, the most accurate model risks propagating these biases in the future, whereas the fairest model risks increased inaccuracies due to a lack of data on how we would like the world to be versus how it actually is. Developing a fair model can be an act of guesswork because although some aspects may be balanced to represent our social ideals, others will reflect the effects of the existing, non-ideal conditions. 

As this section should make poignant, machine-learning algorithms are designed to encode bias into a predictive model that reflects the relevant statistics of the training data. When systemic bias corrupts the data sources, how can anyone hope to build a fair AI system? To stave off pessimism, we note that steps have been taken to address this concern. Companies are developing products, such as IBM's AI Fairness 360 toolkit \parencite{Varshney2018} and Microsoft's Fairlearn \parencite{bird2020fairlearn}, with claims that they evaluate systems for unwanted biases. Even if these products work, the concern remains that the solutions they provide are inadequate. While such tools may enable more thorough testing and monitoring of deployed systems, they stop short of using automated methods to prevent and remediate the propagation of social biases.

Having considered the premises of Syllogism 3 in detail, we are now in a position to present a modified version in which the conclusion is not patently absurd. In this version, the first premise is more specific but fits with sectarian beliefs in AI regarding the importance of machine learning as a foundational technology. Furthermore, the second premise specifies a particular form of bias, opting for the form that typically concerns the general public.
\pagebreak
\begin{quote}
    \textbf{Syllogism 3*}\\
    People are general learning algorithms.\\
    General learning algorithms are not societally biased before being exposed to data.\\
    $\therefore$ People are not societally biased before being exposed to data.
\end{quote}
 With these amendments, the conclusion seems acceptable and at least not immediately dismissible. People, inasmuch as they start out as general learning algorithms presumably cannot be biased before being exposed to experiences in the world. As they are exposed to the world, they are vulnerable to learning the patterns of bias ingrained in society. 

 Even though it is no longer absurd, is the conclusion of Syllogism 3* acceptable? For it to be true, one has to assume that there are no inductive biases (e.g., in genetically determined neural pathways) that favor sorting people into categories and treating those categories differently. To put it more bluntly, a person is not the product of culturally or societally motivated selective pressures that affect the evolution of neural structure or mechanisms. Although evidence indicates that biases around race and gender are measurable in preschool age children \parencite{Perszyk2019}, whether this is due to genetically ingrained, inductive biases; sample biases due to the structure of the child's environment; the insidious influence of societal biases during learning; or other factors is unclear. Suppose one accepts a   blank-slate view at least in terms of the features that support category formation. That is, features where bias is noxious, such as age, race, ethnicity, gender, and sex, are not ingrained. Is it sensible to talk about a person as something dissociable from its life history? Is a model built via machine learning dissociable from the data that informed it? The answers are not trivial. What we can say is that people, unlike essentially all AI systems, have the capacity to consider their learned responses and rebel against their history, against the common practices of their societies. The decisions that we make and the actions that we take, even as children, are potential targets for deliberation and susceptible to veto by reason.

\section{Diagnosis}

Surprisingly (to the authors, at least), discussions among researchers about the potential for bias in AI systems can become heated.\footnote{We are explicitly avoiding finger pointing, but high-energy exchanges on the topic are easily found on the internet.} If these were arguments about whether systems should be evaluated for biased output, then data collection procedures should be available for review and dataset statistics should provide information about sensitive dimensions. Even if the underlying algorithms can be biased in some sense, then why would ordinarily level-headed people become emotionally charged? Our conjecture is that the source of outrage is not dry technical concerns, but instead results from treating AI systems as ersatz moral agents; machines that can sentence criminals, select job applicants to interview, or even determine who is categorized as human. Moreover, we suspect that there are three separate kinds of outrage sparked by these discussions: intellectual, moral, and political.

\subsection{Confusion}
\label{sec:confusion}
Intellectual outrage is a rarefied form that centers on what is treated as a person. Everyone would agree that toasters and rabbits are not people. Even when machines incorporate AI technologies, such as rice cookers that use fuzzy control to steam more reliably and automatic transmissions that use adaptive learning to shift more effectively, no one would declare them to be persons. In that same vein, the algorithms that exhibit bias are not considered people. So, when we claim that there is intellectual outrage, we are not asserting that there is a public argument in which one side asserts that these algorithms are persons, and the other asserts the opposite. Instead, we suspect that both the way these algorithms are used and the capacities ascribed to them implicitly suggest a status of personhood. When computers mimic a person's abilities to reason, decide, and act to produce behavior that surprises us, such as winning games of Go or Jeopardy against human experts, lines appear to blur. Some people may claim that AI systems like AlphaGo Zero \parencite{Silver2017,silver2017mastering,marcus2018innateness} and Watson \parencite{Ferrucci2011} capture the important functionalities of thought and action while others object.

In the current context, the ascription of bias to computer programs, and specifically the sort of social bias discussed in section~\ref{sec:bias}, blurs the lines of personhood. Bias in a sense that supports moral culpability has specific relationships with reasoning, deciding, and acting that none of the AI systems accused of bias encode. For a system to be biased in this way, it must have a variety of capacities:
\begin{enumerate}[label=(\alph*)]
    \item to entertain alternatives using explicit representations,
    \item to apply reasoning to determine which alternatives align with normative preferences,
    \item to use the results of reasoning as a guide to deciding which alternative to select, and
    \item to intentionally apply the chosen alternative through available behavioral routines.
\end{enumerate}
If any of these are missing, a system cannot be biased in the way that a person can. If a system lacks (a), then it has no access to its alternatives and can only operate as a simple function, mapping its input to some output. If a system lacks (b), there is a disconnect between preferences and alternatives such that any ability to enforce or correct for biases is severed from a decision-making process. Without (c), a system can only flip a weighted coin among available alternatives, ceding choice to randomness. And in the absence of (d), any resulting behavior is not a proper action because it is divorced from situational preferences and rational evaluation. So, when people claim that an algorithm is (socially) biased, there is an implication that it is  more like a person than it actually is. People sensitive to the overreach of this claim may become intellectually outraged and feel the need to correct it on technical points even if they are themselves unaware of the source of the outrage.

\subsection{Dehumanization}

Moral outrage is a reaction to the (ironic) dehumanization felt when activities typically thought of as requiring something uniquely human are assigned to computers or other machines. In the United States, the automation boom that started in the 1950s led to fears that massive disemployment would follow as machines replaced skilled labor. The situation regarding AI and automation is different. Even as robots wander around massive warehouses filling online orders alongside, but eventually instead of, humans, the future march of physical automation inspires less outrage than its mental counterparts. The role of people in organizations is shifting from making decisions based on personal expertise to affirming recommendations output by programs. The lives of people are being altered by AI systems that, for all practical purposes, hire employees, sentence criminals, and approve loans. Historically, these kinds of decisions would rely on both the broad context and individual circumstances of a specific case. 

In any discussion of biases in decision-making when there are human stakes, a critical component is that mistakes brought about by ignorance or prejudice can weigh on one's conscience, a capacity that AI systems do not have. We recognize that one perspective on conscience is that it works as a moral guide, helping people decide what is good or bad according to some internalized norms. On this account, the idea of a computational conscience is not out of the question. However, we point out that one's conscience is also often a motivational force, ensuring that we spend time on decisions with moral weight. When we can be judged as good or bad on a moral spectrum for our choices, we generally think longer and more deeply about the consequences of each situation and the norms that apply. From this standpoint, the nature of a computational conscience is less clear. Putting aside whether building this characteristic into an AI system is possible or even necessary, offloading decisions with moral weight (i.e., that ordinarily have subjectively experienced consequences on the deciders) to a machine that lacks subjectivity seems like cheating. 

So, one source of moral outrage is that important decisions that can affect the lives of several people are being made by a machine that has no understanding of goodness, fairness, or justice. The factors contributing to this outrage and the very real effects of AI in society are treated by authors trained in critical theory and the subarea of critical science \parencite{Birhane2021,Noble2018,Mohamed2020}.\footnote{
    In philosophy, critical theorists approach knowledge in a markedly different way from scientists. First, their analyses start with a value-based assumption (e.g., all of human society is structured by class differences), and they interpret various social interactions through the lens of their assumption. Second, the goal of their analyses is emancipatory, either in the sense of maximizing human freedom (e.g., the elimination of class structure in a communist utopia) or in the sense of identifying and pushing back against restrictions masked by prevailing ideology (e.g., an emphasis on rationality, seen as liberating in the Enlightenment period, has led to the formation of new power structures and means of dominance). The current paper takes scientific perspective and uses an inconsistency within that perspective to argue for a broadening of scope in AI research. Readers are encouraged to explore articles by scholars who are applying critical theory to AI, some of whom we have cited, keeping in mind that the epistemology of science and the epistemology of critical theory are at odds. We direct readers interested in learning more about critical theory in general to the Stanford Encyclopedia of Philosophy \parencite{sep-critical-theory}.} Another source is that an artificial decider can make poor, biased choices without suffering the pangs of conscience. A related source is that humans running such AI systems can avoid feelings of guilt and regret by inaptly placing the moral burden on the program. Dehumanization, therefore, is two-pronged. First, it is dehumanizing for a machine to tell a person their worth. Second, it is dehumanizing when people  treat conscience, and relatedly the moral weight of actions, as irrelevant in cases where a person would feel its pressure. When arguments use the language of social bias, they imply a sense of fairness that risks bringing these concerns to the forefront and adding an emotional charge to an otherwise reasoned discussion.

\subsection{Disempowerment}

If dehumanization is one side of the coin, disempowerment is the other. Moral outrage results from the disassociation of activities from the subjective experiences that people consider vital to successful, properly grounded, intentional actions. Political outrage is felt by the people on the receiving end of those actions. When a computer denies your loan and the loan officer says, ``There is nothing I can do,'' to whom do you appeal? The officer is no longer a decision-maker in this situation. Instead, their role as an agent is reduced to the ritual officiant who, through a performative utterance, legally approves or denies the loan. The locus of power has been ceded to a program that, to paraphrase Edward Thurlow, has no body to punish and no soul to condemn \parencite[][p. 268]{Poynder1844}. 

Thurlow made his original statement about corporate entities in the 1700s, and since then, legal systems have developed methods to restrict and punish human organizations. No such rules exist for AI systems, and society has no settled means for restricting or punishing their actions when they produce objectionable output \parencite{Danaher2016}. The most people can do is to seek restitution from or to direct retribution toward whoever is responsible for the use of the systems. The software itself may then be rewritten, retrained, or retired to eliminate unwanted biases. Nevertheless, the existence of the program and its application reveals the extent to which we cede our power to algorithms that have no heart in which to treasure our best interests. To the extent that AI is used to approve loans, to sift resumes, to sentence criminals, and to filter speech, the door is open for political outrage. All that is required is another unjust decision or a shift in mores that produces a misalignment between societal values and those encoded within existing AI tools. 

\section{Prescriptions}

The AI systems built and deployed so far are not moral agents. They cannot determine right from wrong, and they cannot be held accountable for their behaviors. In the previous section, we suggested that placing these systems in roles typically reserved for moral agents (i.e., for people) leads to outrage. The same emotional responses are rightfully not directed at cruise control, mail sorters, assembly-line robots, and other forms of automation. However, we anticipate that not only will AI continue to replace people in these kinds of roles, their encroachment into the territory of genuine moral agents will proceed. Instead of belaboring claims of bias as if it is entirely avoidable, society needs to reconcile itself to these new applications of automation. In our view, there are three steps that can mitigate outrage while taking seriously the challenge of reducing unwanted bias in AI. 

\subsection{Refine the Language}

The etiology of intellectual outrage is in an unjustified anthropomorphism: the ascription of the properties and functions of personhood to entities---AI programs---that are not persons in any sense.\footnote{Bringsjord \parencite*{Bringsjord1992} provides one possible argument against the potential personhood of AI programs.} We suspect that these ascriptions are accidental and due to the use of ambiguous language or general carelessness. AI researchers, the authors included, often talk about their systems as having beliefs and goals, as reasoning and making decisions, and as acting in the world. When speaking this way, these words do not carry the psychological baggage of their ordinary meanings. Even when we claim that beliefs, goals, and plans in a specific system are analogous to beliefs, goals, and plans that people have, the connections are thin---emphasizing limited, shared, functional characteristics. The language around bias, as we have pointed out, is trickier because both people and algorithms have inductive bias, are susceptible to sampling bias, and can exacerbate social biases. However, bias in its deeper, common use is more than subjection to the churnings of unconscious associations or programmed preferences.

In section~\ref{sec:bias}, we claimed that a biased person must have, at a bare minimum, four abilities related to representation, reasoning, decision-making, and acting. Gordon Allport also noted, ``If a person is capable of rectifying his erroneous judgments in the light of new evidence, he is not prejudiced'' \parencite*[][p. 9]{Allport1979}. We take the word `capable' in this quote to include both the ability to correct a revealed partiality and a decision process by which a true correction would be accepted. If either of these conditions are unmet, then someone or something would not be prejudiced. In people, prejudice typically asserts itself as a stubbornness, a resistence to fact. There is rarely a question regarding the ability to change one's views in principle. Contrast this with existing AI systems, even those that propagate social bias, which cannot even rectify their judgments without a person stepping in to reprogram or retrain the models. To be clear, our argument is not that we should consider bringing about actually existing, prejudiced AI---quite the contrary. Instead, we would reserve \textit{bias} in its various forms as a technical term applicable to persons and non-persons alike and use the word \textit{prejudice} to refer to those characteristics of bias implied by personhood and all the complexities of such an implication.

\subsection{Police the Systems}

Language changes may improve the emotional tenor of intellectual discussions, but preventing AI from exacerbating unwanted social biases requires active policing. One suggestion is to audit systems for social bias \parencite{Raji2020}. These audits can be internal, typically organized and funded by the company developing the system, or external, run by a third-party. Although external audits can garner public trust, they are limited by the accessibility of the code, data, and procedures used to develop the system which are typically protected intellectual property. Consequently, there has been an interest in cooperative audits such as one arranged between the company Pymetrics and researchers at Northeastern \parencite{Wilson2021}. The methods developed for that effort attempted to balance the need for the company to protect its trade secrets with the third-party's need to have enough access to make meaningful and responsible claims about Pymetrics' job-candidate recommendation tools. Of recent note, New York City Local Law 2021/144 was codified in December of 2021 to regulate "automated employment decision tools" and is a significant step toward auditing AI bias in one domain \parencite{Carter2022}. More recently, The Office of Science and Technology Policy \parencite*{ostp2022} has published its ``Blueprint for an AI Bill of Rights,'' which outlines a position on protections against discrimination by automated systems. 

Even if audits were to become typical, timely, and trustworthy, there are other concerns that they cannot address. One issue is that audits take time and effort. This sort of delay may be acceptable in some cases, but in others we would prefer to see self-monitoring, self-correcting tools. This is especially the case when the application of a system shifts practices in a way that alters the distribution of future data. Monitoring for effects in real time, especially mismatches between the expected characteristics of a dataset and the observed characteristics, could enable the detection of unintended consequences before an appropriate audit could take place. AI systems are necessarily built and validated using data derived from patterns of behavior established before they were created. The effects of deployment must, therefore, be under keen scrutiny. Shifts in underlying data distributions should be automatically monitored and, if necessary, the AI systems should be realigned and redeployed. 

\subsection{Design for Rebellion}

The final step that we suggest may sound more like science fiction than fact, but researchers have raised the possibility of building rebel agents \parencite{Coman2018,Briggs2015}, AI systems that can refuse to execute plans, pursue goals, and follow orders. The rough idea is that executing some orders may result in harm unforeseeable when they were given. In such a situation, the AI agent should refuse to comply with the order and be able to provide a rationale for its non-compliance. The sort of rebellion we have in mind is internal. In section~\ref{sec:systemic-bias}, we pointed out that people have the ability to rebel against their own learned responses, to explicitly veto courses of action that violate norms. A rebel-agent framework could indicate how to build tools that operate in an analogous way.

Consider a two-tiered AI system that, at the first level, produces ranked alternatives which reflect the statistics of its training data. At a second level, these alternatives are evaluated against a set of norms and are subjected to other forms of monitoring to ensure that standards of fairness are met. This design would enable capacities a, b, and c introduced in section~\ref{sec:confusion}, which are required to exhibit bias in a human-like manner. The value here is that the biases encoded in the norms could be explicitly represented and interpretable and would form a barrier of protection against the inductive and/or sampling biases that would otherwise lead to socially biased activity. 

\section{Conclusion}

We began our investigation of bias in machine learning and AI with Syllogism 3, which helped us reveal underlying concerns. At the basic level, these concerns are about the use of objects to do the sort of activities traditionally assigned to humans. Looking more closely, it seems that people are unconvinced that AI systems have the capacities that these activities require, such as the abilities to self monitor, to self-correct, and to act conscientiously. As we continue to research and develop systems to fill these roles, we may find that the capacities seen as unique to humans were not as necessary as we thought. Alternatively, we may identify the functional roles that these capacities play and determine how to implement them computationally. 

Although this paper has been focused on issues of bias, the general result is applicable to other areas in the purview of AI. Just as algorithms are not biased in the way that people are, they do not have cognitive states in the way that people do. Beliefs, desires, intentions, and other mental attitudes remain uniquely tied to human agents. As we have discussed in the context of AI bias,  programs lack many of the functions that these attitudes afford people. Whenever we implicitly or explicitly ascribe such mental attitudes to computer programs, we risk misunderstandings in the public square. These misunderstandings can turn into strong disagreements and heated arguments which, in turn, interfere with work to bring AI technology nearer to adequately filling human roles. In these and similar cases, we should continue to pursue clarity in our language, vigilance in testing AI software, and boldness in designing increasingly capable systems.

\section*{Acknowledgements}
This paper has benefitted from the suggestions of Anthony Beavers, Marcello Guarini, Alistair Isaac, Patrick Lin, and Matthias Scheutz. The work of authors WB and PB on this paper was supported by the Office of Naval Research award number N0001422WX00138. Past support that enabled author SB with collaborators to invent, specify, and test person-\textit{like} ethical reasoning and decision-making in artificial agents/robots was provided by the Office of Naval Research through a Multi-University Research Initiative Award entitled Moral Competence in Computational Architectures for Robots (PI Matthias Scheutz, Co-PIs Bertram Malle, Selmer Bringsjord). 
Distribution Statement A. Approved for public release; distribution is unlimited.

\printbibliography
\end{document}